\documentclass[prx,twocolumn]{revtex4-2}
\usepackage[x11names, rgb, html]{xcolor}

\definecolor{red}{HTML}{f54b1a}
\definecolor{pink}{HTML}{d19eb1}
\definecolor{orange}{HTML}{d3772e}
\definecolor{yellow}{HTML}{ebe85d}
\definecolor{green}{HTML}{0f6852}
\definecolor{lightblue}{HTML}{01abe9}
\definecolor{darkblue}{HTML}{1b346c}
\definecolor{tan}{HTML}{e5c39e}
\definecolor{darktan}{HTML}{af9e73}
\definecolor{grey}{HTML}{c3ced0}
\definecolor{darkgrey}{HTML}{9dadc4}
\definecolor{black}{HTML}{110d1b}
\definecolor{white}{HTML}{f1f8f1}

\usepackage{hyperref}
\hypersetup{
  colorlinks = true,
  linkcolor = red,
  citecolor = darkblue,
  urlcolor = lightblue
}

\usepackage[cmintegrals,cmbraces]{newtxmath}

\usepackage{graphicx}

\usepackage[margin=1in]{geometry}

\usepackage{algorithm}
\usepackage[noend]{algpseudocode}
\algrenewcommand{\algorithmiccomment}[1]{$\vartriangleright$ #1}
\algrenewcommand{\algorithmicreturn}{\textbf{Return: }}
\algnewcommand\algorithmicinput{\textbf{Input: }}
\algnewcommand\Input{\State \algorithmicinput}

\def\ab{\boldsymbol{a}}

\def\eb{\boldsymbol{e}}

\def\sb{\boldsymbol{s}}
\def\ub{\boldsymbol{u}}

\def\xb{\boldsymbol{x}}

\def\Fb{\boldsymbol{F}}

\def\Xb{\boldsymbol{X}}

\def\EE{\mathbb{E}}

\newcommand{\avg}[1]{\left\langle #1 \right\rangle}

\def\<{\langle} \def\>{\rangle}

\DeclareRobustCommand{\argmin}{\operatorname*{argmin}}

\usepackage{dcolumn}
\usepackage{bm}
\usepackage{float}

\begin{document}
\title{Adaptive nonequilibrium design of actin-based metamaterials: fundamental and practical limits of control}
\author{Shriram Chennakesavalu}
\author{Sreekanth K. Manikandan}
\author{Frank Hu}
\author{Grant M. Rotskoff}
\email{rotskoff@stanford.edu}
\affiliation{Department of Chemistry, Stanford University, Stanford, CA, USA 94305}
\date{\today}
\date{\today}
\begin{abstract}
   The adaptive and surprising emergent properties of biological materials self-assembled in far-from-equilibrium environments serve as an inspiration for efforts to design nanomaterials and their properties.
   In particular, controlling the conditions of self-assembly can modulate material properties, but there is no systematic understanding of either how to parameterize this control or how \emph{controllable} a given material can be. 
   Here, we demonstrate that branched actin networks can be encoded with \textit{metamaterial} properties by dynamically controlling the applied force under which they grow, and that the protocols can be selected using multi-task reinforcement learning.
   These actin networks have tunable responses over a large dynamic range depending on the chosen external protocol, providing a pathway to encoding ``memory'' within these structures. 
   Interestingly, we show that encoding memory requires dissipation and the rate of encoding is constrained by the flow of entropy---both physical and information theoretical.
    Taken together, these results emphasize the utility and necessity of nonequilibrium control for designing self-assembled nanostructures.
\end{abstract}

\maketitle
\section{Introduction}

Biomolecular materials spontaneously self-assemble in fluctuating, nonequilibrium environments.
Despite this complex environment, many cellular assemblies possess a remarkable capacity for adaptation and dynamism~\cite{pollard_cellular_2003,fletcher_active_2009,janmey_cytoskeleton_1998,janmey_cell_2007}.
In contrast, our ability to design materials with nanoscale components remains primitive in comparison with the intricate, hierarchical structures that reliably assemble in living systems~\cite{whitelam_statistical_2015}. 
The strategies employed by biological self-assembly thus offer a unique lens into the limits of control for synthetic efforts to engineer nanoscale metamaterials. 

Cytoskeletal networks are perhaps the most illustrative example of the dramatic effects of assembly dynamics on resulting material properties~\cite{pollard_regulation_2007,mackintosh_elasticity_1995}.
The key structural proteins that form this scaffolding of the cell must rapidly rearrange to drive fundamental processes such as motility~\cite{gardel_mechanical_2010} and signalling~\cite{janmey_cytoskeleton_1998,pollard_actin_2009}. 
The remarkable material properties and dynamical assembly of branched actin networks, which self-organize through the complex interplay between numerous proteins and external forces from the environment, have attracted significant experimental and theoretical scrutiny.
The components of these dynamic networks have been purified, and reconstituted systems have been studied extensively~\cite{bieling_force_2016,parekh_loading_2005,shaevitz_load_2007,li_molecular_2022,mackintosh_elasticity_1995}, yielding detailed insight into their nonequilibrium dynamics.

Atomistic simulations and experimental studies have provided insight into the molecular mechanisms of actin polymerization and allostery~\cite{chu_allostery_2005}, branching~\cite{ding_structure_2022}, and plasticity~\cite{schramm_plastic_2019}.
\emph{In silico} studies of mesoscale material properties of cytoskeletal networks have, on the other hand, relied heavily on minimal models~\cite{popov_medyan_2016,freedman_versatile_2017,mackintosh_elasticity_1995}, that represent the network of actin filaments as elastic networks with filaments that resist bending and stretching.
For example, randomly assembled networks have been shown to recapitulate interaction with the membrane~\cite{liu_membraneinduced_2008} and myosin induced contraction~\cite{popov_medyan_2016,freedman_versatile_2017}.
Similarly, interesting dynamical behaviors like treadmilling~\cite{hohlfeld_communication_2014} as well as aster formation and sorting in actin-myosin networks~\cite{lamtyugina_thermodynamic_2022,qiu_strong_2021} have been captured accurately with simple models.
The aforementioned works show the utility of a minimal modeling approach, and we follow suit. We develop a model that closely resembles the experimental geometry of a recent study that images branched actin network growth under load~\cite{li_molecular_2022,bieling_force_2016}.

We show here that the model we have developed captures the essential behavior of actin network growth under varying external loads and, additionally, we show how to control that behavior to tune the response properties of the resulting material. We demonstrate, in particular, that an external feedback protocol that dynamically modulates the growth conditions encodes a nonequilibrium memory in the material. 
The deviation from a homogeneous distribution of the molecular components leads to distinctive metamaterial properties, but reducing the system entropy relative to the homogeneous state comes with an associated entropic cost. 
Using reinforcement learning, we optimize external, nonequilibrium protocols that drive the resulting networks into regimes inaccessible to uncontrolled assemblies.
We further show that the resulting networks have elastic coefficients that depend strongly on the applied force.

\section{Encoding material memory with nonequilibrium growth dynamics}
\begin{figure*}
    \centering
    \includegraphics[width=1.0\linewidth]{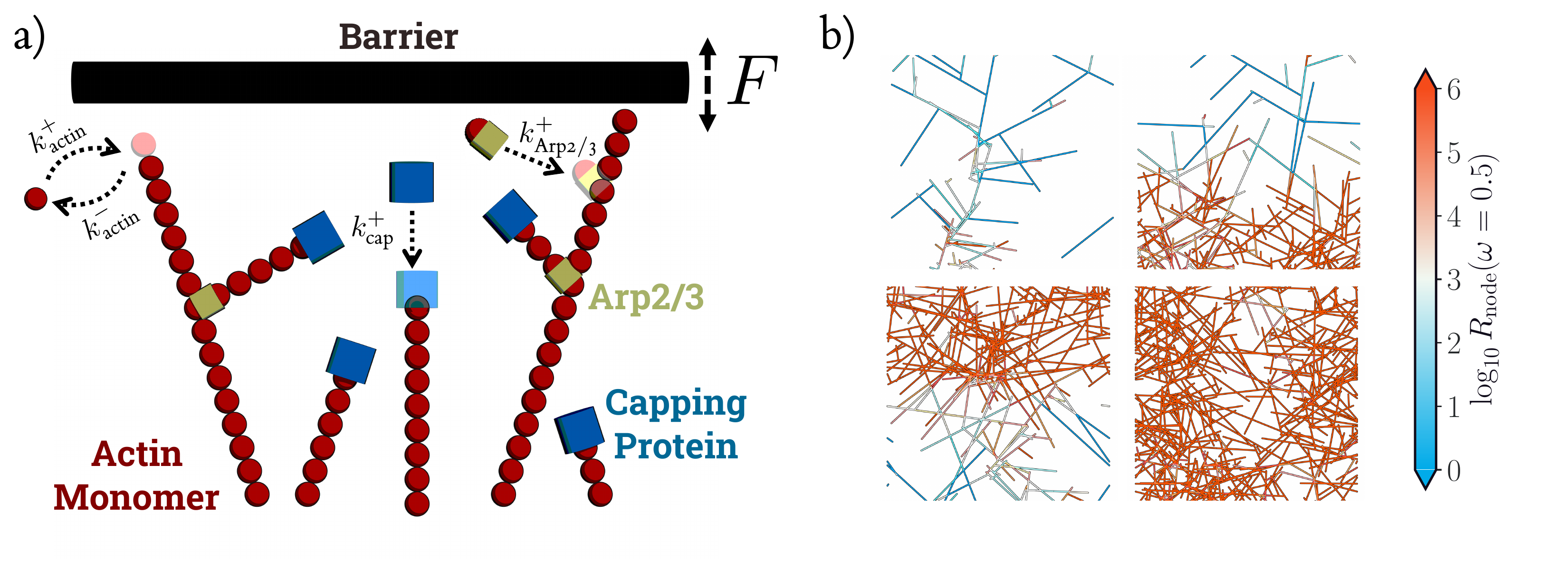}
    \caption{External control of actin network enables design of actin-based metamaterials. a) A schematic overview of branched actin networks growing against a barrier, where individual filaments can polymerize, depolymerize, branch (via Arp2/3) or be capped. Network growth can be controlled by modulating load force $F$. b) 
    Actin networks grown with layered density profiles (top right and bottom left) have
    heterogeneous responses, while networks grown with constant density profiles (top left and bottom right) have more homogenous responses.}
    \label{fig:schematic}
\end{figure*}

Mechanical response in soft materials is dictated by both intrinsic material properties of the constituents and also their spatial organization~\cite{ronellenfitsch_inverse_2019,rocks_designing_2017}.
When the individual components of a structure can be manipulated in a spatially localized fashion, exquisitely precise control of the resulting material is sometimes possible~\cite{ronellenfitsch_inverse_2019,falk_learning_2022}.
However, in many materials, we can only reconfigure a fixed set of components into distinct topologies. This motivates an exploration of the possibility of controlling mechanical response without manipulating the components themselves.
For example, the reorganization of fixed components in a network has been used to control global response by systematically removing a small set of bonds from disordered networks of springs~\cite{goodrich_principle_2015}.

Making metamaterials by manipulating a few degrees of freedom in a complex network indeed demonstrates the profound sensitivity of mechanical properties on spatial organization.
However, in nanoscale systems self-assembled by stochastic dynamics, surgical reconfiguration is not possible experimentally.
Nevertheless, many biomolecular materials have adaptive properties that evince \emph{memory} of their growth conditions~\cite{discher_tissue_2005,gardel_prestressed_2006,janmey_cell_2007},  which are ultimately nonequilibrium, kinetic effects.
Perhaps the most direct example of this behavior is the force-induced stress stiffening in branched actin networks~\cite{shaevitz_load_2007,parekh_loading_2005,storm_nonlinear_2005} which can be reversibly saturated to the point of softening~\cite{chaudhuri_reversible_2007}.
When grown under high load forces, actin networks become more rigid than networks assembled in the absence of an applied force, despite consisting of exactly the same protein components. 

The metamaterials that assemble under the external control described in Sec.~\ref{sec:directed} have a ``memory'' of their growth process.
In many works examining nonequilibrium memory retrieval and memory in self-assembly, the existence of a Hamiltonian with multiple metastable states is assumed~\cite{zhong_associative_2017}; here we instead examine the question of the thermodynamic constraints on \textit{encoding} memory and generalize this notion to non-Hamiltonian systems. 
This memory can be quantified by considering the deviation of the spatial distribution of components after the finite-time protocol $\rho_T$ from a reference homogeneous steady state distribution, denoted $\rho_{\rm ss}$. 
A more structured distribution will have lower Shannon entropy $S_{\rm sys}(\rho)\equiv -\int \rho(\xb) \log \rho(\xb) d\xb,$ setting $k_{\rm B}=1$, so a time-dependent reduction of the system entropy constitutes encoding memory.

Interestingly, the rate of memory encoding satisfies a universal ``speed-limit'' as a direct and straightforward consequence of the detailed Information Fluctuation Theorem~\cite{parrondo_thermodynamics_2015,horowitz_nonequilibrium_2010a}:
\begin{equation}
    -\partial_t S_{\rm sys}(\rho_T) \leq  \avg{ \sigma_{\rm env}} + \dot I[\Xb_t; \mathcal{M}_t],
    \label{eq:speedlimit}
\end{equation} 
where $\sigma_{\rm env}$ is the rate of entropy production in the medium and the brackets $\avg{\cdot}$ denote averages over trajectories.
To see this, we employ the detailed fluctuation theorem assuming that the material is growing under steady state conditions at time $t=0$, which identifies the total entropy production along a feedback controlled trajectory as 
\begin{equation}
    \omega[\Xb_t, \mathcal{M}_t] = \log \frac{\rho_{\rm ss}(\Xb_0)}{\rho_{T}(\Xb_T)} - \beta \mathcal{Q}[\Xb_t, \mathcal{M}_t] + \mathcal{I}[\Xb_t; \mathcal{M}_t].
\end{equation}
Due to the second law of thermodynamics, this quantity is non-negative in expectation, i.e., 
\begin{equation}
    \begin{aligned}
            0 &\leq \EE_{\Xb_t,\mathcal{M}_t} \log \frac{\rho_{\rm ss}(\Xb_0)}{\rho_{T}(\Xb_T)} - \beta \mathcal{Q}[\Xb_t, \mathcal{M}_t] + \mathcal{I}[\Xb_t; \mathcal{M}_t] \\
            &\leq \Delta S_{\rm sys}(\rho_T, \rho_{\rm ss}) - \beta \avg{\mathcal{Q}} + I[\Xb_t; \mathcal{M}_t],
    \end{aligned}
\end{equation}
where the entropy term appears because the initial conditions are sampled from the steady-state distribution which differs from the final distribution and the average of the trajectory-wise information is the mutual information between the trajectory and the measurement sequence.
This term vanishes when there is no feedback in the protocol. 
Because the total heat flow out of the system is time extensive, it is more useful to consider the time derivative of this inequality, which recovers~\eqref{eq:speedlimit}, using the fact that the steady state system entropy change is zero.

The interpretation of this speed-limit is straightforward: the rate at which the system can be driven to deviate from its steady-state behavior requires either a positive rate of dissipation of heat to the reservoir or a positive quantity of information must be extracted from the measurement sequence at each point in time.
The entropy production rate for actin networks grown under feedback protocols is plotted as a function of protocol duration in Fig.~\ref{fig:epr}.
In the special case of an equilibrium reference steady-state, this bound can be saturated because there is no steady-state entropy production; note that if the reference process is dissipative in the steady state, this upper bound may not be tight. 
Bounds similar to~\eqref{eq:speedlimit} have appeared without the feedback term to quantify the cost to maintain a steady state~\cite{horowitz_minimum_2017,chennakesavalu_probing_2021} and as a constraint on number fluctuations in self-assembled lattice gases~\cite{nguyen_design_2016}.
While this upper bound may provide only a weak constraint in some cases, it emphasizes that finite-time memory encoding \emph{requires} dissipation.

The general thermodynamic constraint~\eqref{eq:speedlimit} implies that a finite dissipation rate is required to encode new properties in a material. 
Indeed, the large range of elasticities observed in branched actin networks that result only from changing the growth conditions raises an important general question: How tuneable is response with a given material assuming that a dissipative, feedback protocol is used to control assembly?
Here, we seek to answer this question using a minimal model of branched actin networks, described in detail in Sec.~\ref{sec:model}.
This model provides an attractive platform for our investigation because this system can be reconstituted using purified proteins and the growth conditions can be directly manipulated using the cantilever arm of an atomic force microscope, as depicted in Fig.~\ref{fig:schematic}.
Our model captures the essential physical features of recent experiments reported in Ref.~\cite{li_molecular_2022} and, because they work with a fixed set of pure protein components, ambiguities related to cellular regulation of protein expression and other nontrivial side effects of biology can be systematically excluded, allowing us to focus on the underlying physical mechanisms of control. 

Obtaining a metamaterial by modulating growth conditions requires external control, and the primary experimentally accessible control variable in this system is the applied force as a function of time.
Throughout, we consider a setting where the controller interacts with the material, allowing both measurement and feedback, which is a feature of the biological systems and could also be tractable experimentally with appropriate microscopy.
Directly tuning elastic coefficients with a feedback protocol is not straightforward because this global property depends on the network structure in a complicated and nonlinear fashion.
External growth forces do, however, directly impact the local density at the growth front, which in turn is strongly connected to the response. 
Throughout, we study the emergent properties of a fixed, final network at the end of its growth trajectory.

\section{Minimal model of actin growth under external load}\label{sec:model}

\begin{figure}[h]
\centering
\includegraphics[width=1.0\linewidth]{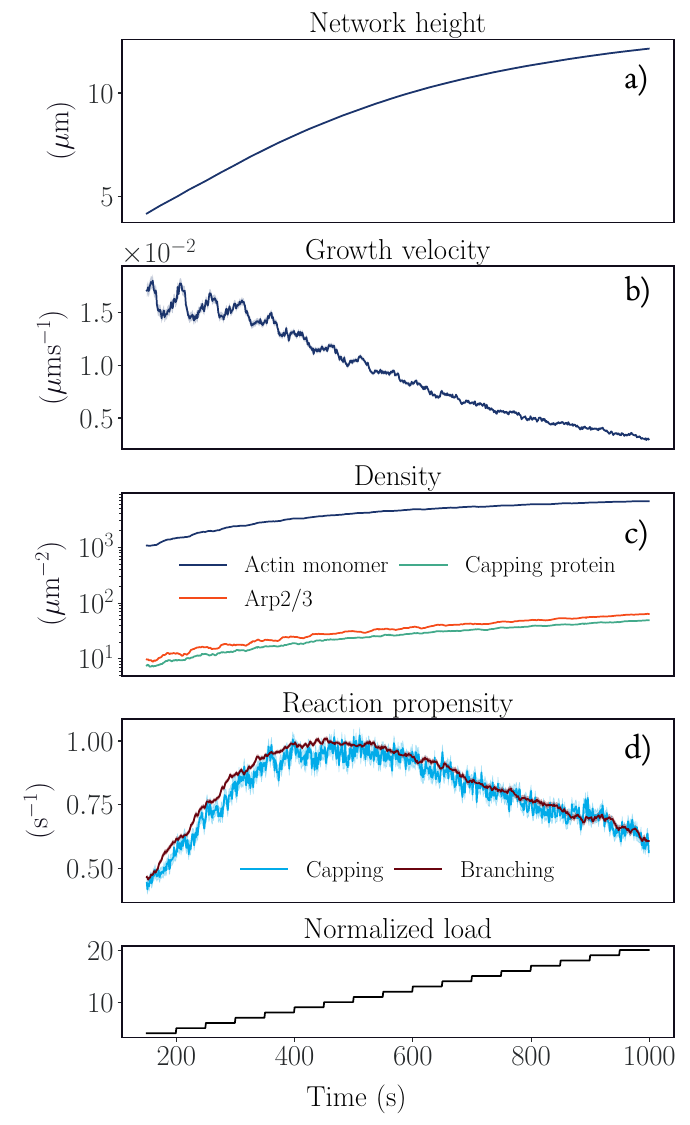}
\caption{Average network height, growth velocity, densities of constituent materials, and reaction propensities for 200 actin networks grown under increasing growth loads (normalized by $f_0$). Under increasing loads (bottom panel), a) network height increases with b) a reduced growth velocity, c) density of actin monomers, capping protein and Arp2/3 within 100 nm}
\label{fig:expts}
\end{figure}

A Brownian ratchet has long been speculated as the mechanism through which external force impacts growth in actin networks~\cite{peskin_cellular_1993, mogilner_cell_1996,mogilner_force_2003,mogilner_polymer_2003,hohlfeld_communication_2014}.
The experiments of Li et al.~\cite{li_molecular_2022} lend additional support to the Brownian ratchet model by imaging components of the assembly during growth.
This set of experiments uses the cantilever arm of an atomic force microscope to precisely apply force to the network in the direction of its growth.
A similar experimental setup was previously used to elucidate stress-stiffening and stress-softening in networks growing under large loads and stress-dependent growth dynamics~\cite{chaudhuri_reversible_2007,parekh_loading_2005, bieling_force_2016}.
These experiments provide fundamental insight into the nature of branched actin network assembly, and in particular, they highlight the delicate balance between 
polymerization, capping and additional branching near nucleation promoting factors.
Additionally, these reconstitution experiments highlight the strong dependence of material properties on growth conditions, even in the absence of biological stimuli.

Because these experiments contain precisely controlled sets of ingredients, we sought to assess if the experimental findings were consistent with a minimalist model of polymerization against a ratchet-like load.
The model we develop is partially inspired by and shares many features with those in Refs.~\cite{maly_selforganization_2001, schaus_selforganization_2007,weichsel_two_2010}.
These two-dimensional models provide useful qualitative insight into the dynamics of self-organization in branched actin networks in the lamellapodium. 
The network growth involves the interplay between ATP-dependent polymerization, Arp2/3-mediated branching of the network, and quasi-irreversible capping of the growing barbed-ends of individual filaments. 
These reactions all have experimentally measured rates, and the physical properties of individual actin filaments have been extensively characterized at the single molecule level (see Table~\ref{tab:actin_rates}).
Remarkably, a Gillespie model that incorporates only these experimentally known parameters predicts emergent, network-level properties that are consistent with experimental results.
While the experimental analogue of the actin networks we consider are three-dimensional, we find, as shown below, that a two-dimensional model is sufficient to explain the experimentally observed behavior.

Our model geometry closely mimics the experimental setup described in Ref.~\cite{li_molecular_2022}.
As depicted in Fig.~\ref{fig:schematic}, actin elongation occurs homogeneously throughout the network with a rate $k_{\rm actin}^+$.
Biologically, actin growth is biased towards the plus end~\cite{janmey_cytoskeleton_1998}, but the experimental setup that we compare most directly with includes only profilin-bound actin which exclusively binds the plus end; that is, no minus end growth occurs in our model. 
Polymerization is balanced by a reverse reaction in which actin spontaneously unbinds from existing filaments with a rate $k_{\rm actin}^-.$
The imbalance in the forward and reverse growth rates leads to a steady-state positive growth velocity. 
For actin and all other components of the network, we fix the chemical potential and assume there is sufficient excess of protein that depletion effects can be neglected. 

Near the barrier, nucleation promoting factors encourage branching of actin filaments via binding of the protein complex Arp2/3~\cite{li_molecular_2022}.
We do not explicitly represent nucleation promoting factors, but capture this proximity effect by allowing Arp2/3 to bind to filaments within $10\ell_{\rm actin}$ of the barrier, where $\ell_{\rm actin}$ denotes the diameter of an actin monomer.
Interactions between the growing end of an actin filament and nucleation promoting factors can disrupt branching~\cite{li_molecular_2022}, so we mollify the rate of branching based on the barbed end density near the growth front.
Explicitly, the branching propensity is given by
\begin{align}
    \begin{split}
    r_{\rm branching} = k_{\rm Arp2/3}^+  n_{\rm{Arp2/3\ binding\ sites}}&\\ \times\ n_{\rm{eff\ Arp2/3}}  [\rm{Arp2/3\ Monomers}]&,
    \end{split}
\end{align}
where $n_{\rm free}$ is the number of free barbed ends, 
\begin{equation}
    n_{\rm{eff\ Arp2/3}} = \max\bigl(0, n_{\rm NPF} - n_{\rm free}\bigr),
\end{equation}
and $n_{\rm NPF}$ is a fixed constant.

The elongation of an individual actin filament can be halted by the so-called capping protein, which binds free barbed ends with high affinity. 
When a filament binds capping protein, an event that is considered irreversible in our model, it no longer polymerizes or depolymerizes.
The only constraint on capping, which occurs homogeneously throughout the network, is steric hindrance with the barrier.


Indeed, sterics dictate the overall growth trajectory of the network via the constraint imposed by the rigid barrier depicted in Fig.~\ref{fig:schematic}.
The Brownian ratchet picture of actin growth under load asserts that polymerization spontaneously occurs as the barrier fluctuates away from the growth front.
In our model, diffusion of the barrier under an applied load is reflected at the point of maximum height of the network.
The transition probability density for such a reflected diffusion can be solved analytically~\eqref{eq:fp_reflective}, meaning that we can propagate the position of the barrier simply by sampling conditionally on time after a reaction occurs. 

Fig.~\ref{fig:expts} illustrates the model branched actin network growth under a step-wise increasing load force, which is in reliable qualitative agreement with experimental observations \cite{li_molecular_2022}. First, under an increasing load, the growth velocity of the actin network decreases. This is primarily a consequence of the increased steric hindrance between the barrier; under higher loads, fluctuations in the barrier that enable polymerization are less likely. Second, the density of the materials near the growth front increases. To enable comparisons with experimental observables, we consider a region of interest within 100 nm of the growth-front, a region comparable to the imaging depth afforded by Total Internal Fluorescent Microscopy (TIRF) \cite{fish_total_2022}. The reduction in growth velocity of the network under an increased load will result in a larger number of uncapped filaments near the growth front of the network, inducing further capping and branching events, resulting in increased densities of actin monomers, capping proteins, and Arp2/3 complexes within this region. Third, we observe that the reaction propensities for branching and capping are balanced across different loads, with both propensities eventually decreasing under an increased load. In this setting, the increased number of free barbed ends interferes with branching events, while the increase in steric hindrance restricts capping of filaments.

\section{Directing self-assembly of actin structures}\label{sec:directed}

\begin{figure}
    \centering
    \includegraphics[width=0.97\linewidth]{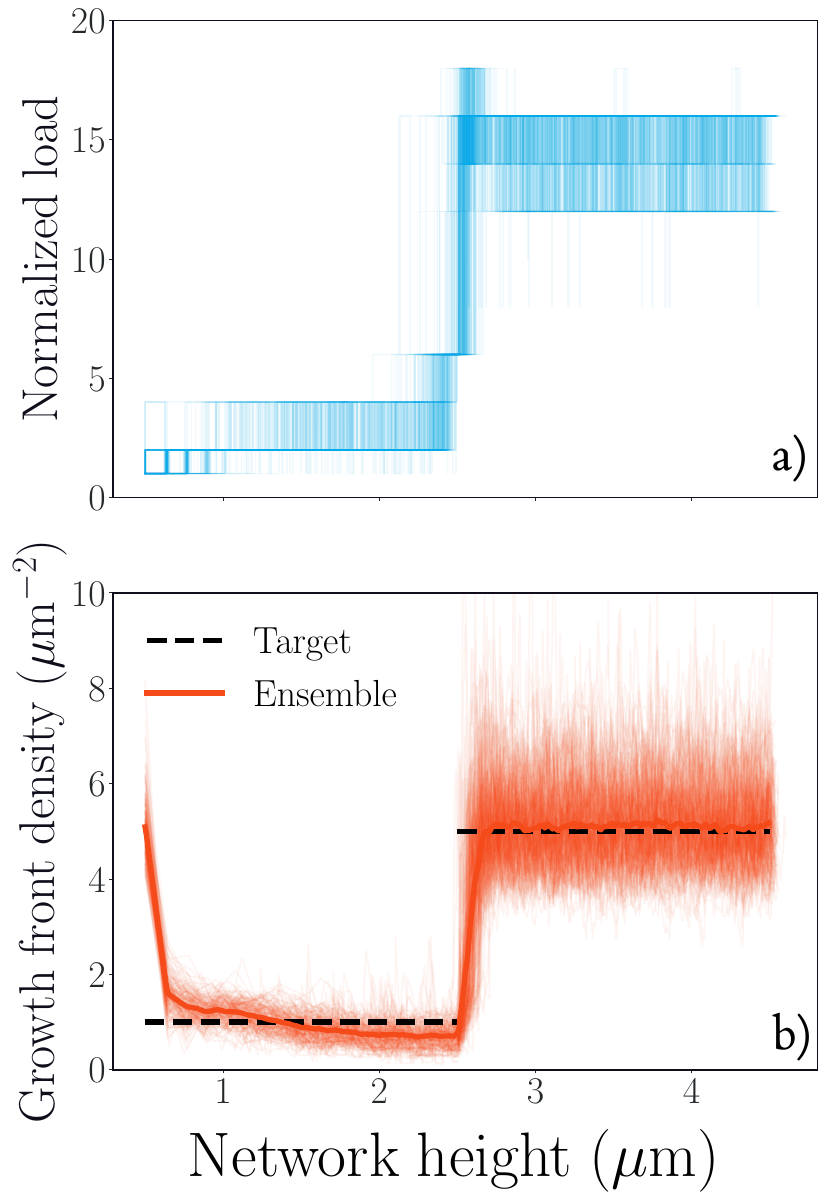}
    \caption{Protocols and growth front density as a function of time for 200 networks grown to a target layered structure with feedback protocols. The ensemble of state-dependent control protocols (a) enable realization of target growth-front density (b).}
    \label{fig:RL_prot}
\end{figure}

The load-induced stiffening by the external load force on a growing actin network suggests the possibility of engineering actin-based metamaterials. 
The spatial organization of the actin network depends strongly on the load history~\cite{parekh_loading_2005}, and furthermore, there is a strong coupling between the response of a network and its spatial organization. 
Previous work \cite{chennakesavalu_probing_2021} has shown that reinforcement learning (RL) offers a robust, yet straightforward framework for nonequilibrium control, when the tunable parameters are extrinsic, such as experimental control variables. 
Following the approach detailed in Ref.~\cite{chennakesavalu_probing_2021}, we devise a multi-task RL strategy (see Appendix~\ref{app:rl}) which allows us to spatially tune the network densities and elicit material responses distinct from those available to homogeneous networks.

The response of actin networks grown under a fixed protocol exhibits high variance, limiting protocol-optimization strategies that directly seek to target a particular response. However, determining protocols that control the density of the network allows us to assemble actin networks in a targeted way because the response of a network is tightly coupled to density.
In practice, we can only reasonably control the growth front of the actin network as filaments that are not proximal to the barrier are quickly capped, and filament branching requires a nucleation promoting factor, which are implicitly present on the barrier. Therefore, we can only regulate branching events on monomers proximal to the barrier. With this in mind, we use RL to design external control protocols that produce specific growth front densities of a network. 
Here, we define the growth-front density as the density of actin monomers within $10\ell_{\rm actin}$ of the barrier.

We consider a discrete set of external loads (see Table~\ref{tab:dql_hyp}) to limit potential experimental challenges of applying load forces with arbitrary resolution. We then train external controllers that can, in a feedback fashion, modulate the external load to achieve a desired density. In Figure~\ref{fig:RL_prot}, we detail trajectories of protocols to grow layered networks with a low-density, soft, foundation and a high-density, stiff, growth-front. The protocols to grow target layered structures are nondeterministic; depending on the current state of the system, different loads are required to maintain a particular density and to transition between different densities. Importantly, we are able to target network densities with high fidelity.

\begin{figure}
    \centering
    \includegraphics[width=\linewidth]{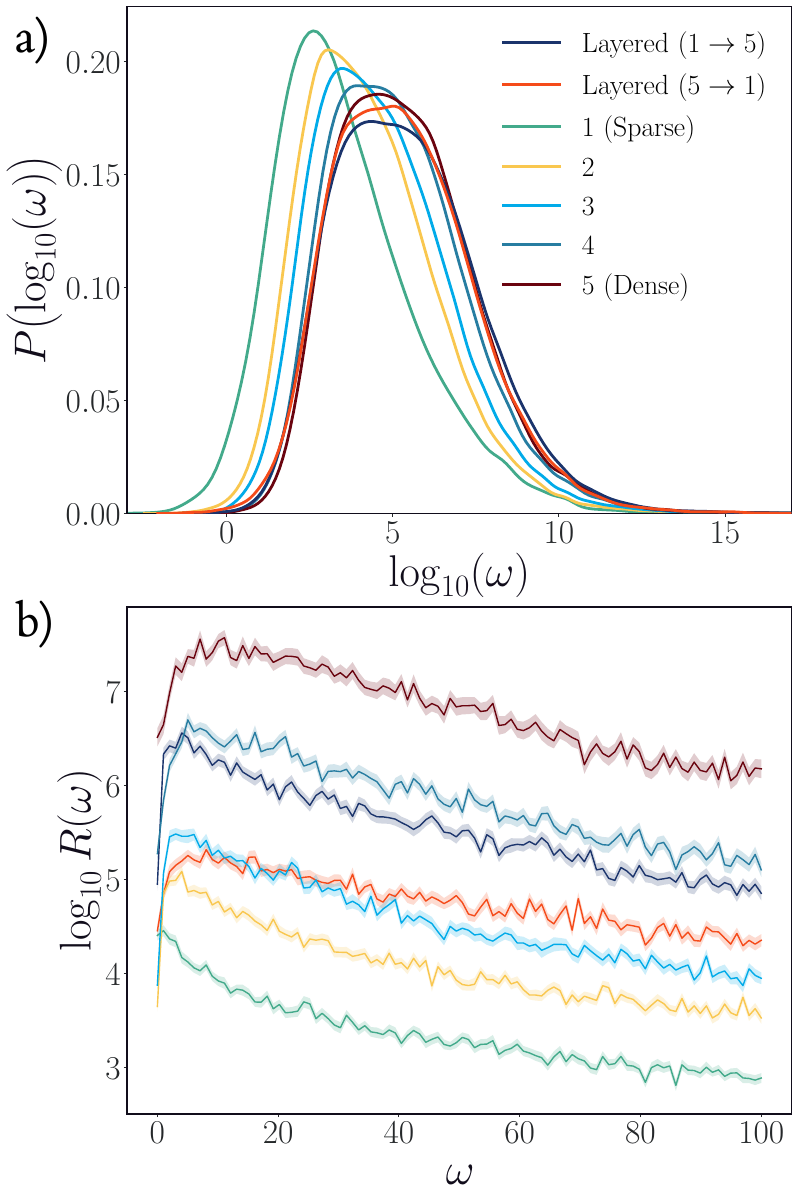}
    \caption{Actin networks targeted to different densities demonstrate diverse responses. 200 networks were grown to target growth-front monomeric densities (1-5) and layered networks were grown with two layers of different growth-front densities ($1\to5$ and $5\to1$). a) The distribution of the eigenvalues for the mass-normalized stiffness matrix (\ref{eq:glob_stiff}, \ref{eq:glob_mass}) is right-shifted under increasing density, with layered densities displaying intermediate distributions. b) The response function---normalized by the number of nodes---$R(\omega)$ \eqref{eq:avg_repsonse} is higher under increasing density with layered densities displaying intermediate responses.}
    \label{fig:response}
\end{figure}

Next, we consider the response properties of actin networks to time-periodic forces. 
We grow networks targeted to different densities, both uniform and layered (Figure~\ref{fig:response}).
For uniform networks, we observe that the distribution of eigenvalues of the mass-normalized stiffness matrix (see Appendix~\ref{app:time_periodic}) is right-shifted under increasing density. Actin networks grown to a higher density will have a larger number of smaller filaments, resulting in higher natural frequencies of the network in comparison to networks with lower density. We further compute the corresponding response functions $R(\omega)$ (see Appendix~\ref{app:time_periodic}), which we normalize by the number of nodes. 
Similarly, for uniform networks, we observe that the denser networks have a higher response as perturbations can more easily be distributed.

\begin{figure}
    \centering
    \includegraphics[width=\linewidth]{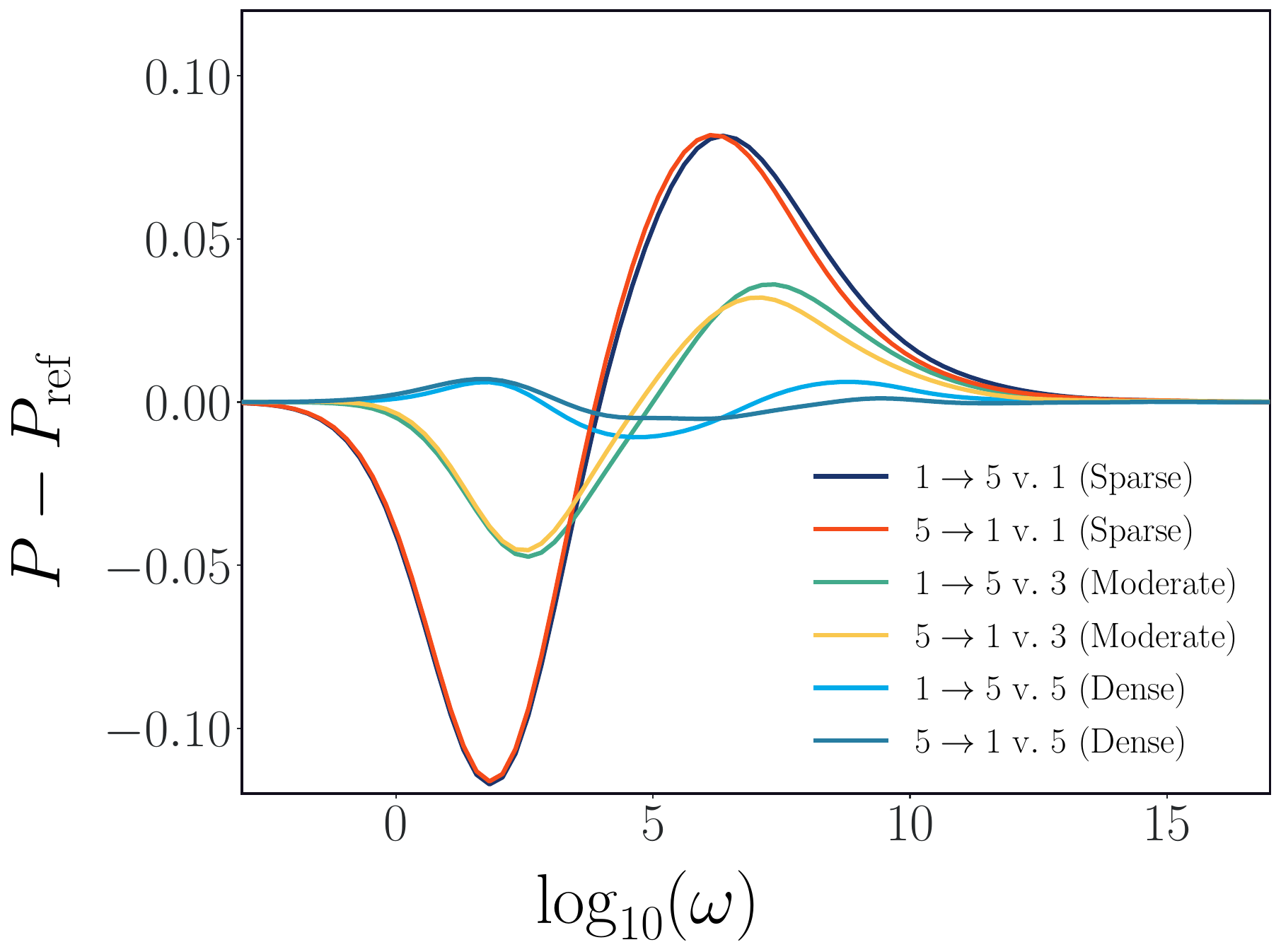}
    \caption{Differences in the density of states between inhomogeneous networks and homogeneous networks. The layered networks respond to perturbations with a profile distinct from those of any homogeneous network. Compared to the ensemble of sparse networks, the layered networks have a suppressed response at low frequencies and an enhanced response at higher frequencies. Compared to the ensemble of dense networks, the layered networks have an enhanced response at low frequencies, and a suppressed response at higher frequencies. These plots indicate the layer networks mix, albeit in a complicated way, the properties of the low density and high density actin networks.}
    \label{fig:enter-label}
\end{figure}

Interestingly, we can engineer spatially tuned responses by growing networks with heterogeneous density profiles. 
We target networks with a low-density foundation and a high-density growth-front (1$\to$ 5) and a high-density foundation with a low-density growth-front (5$\to$ 1). 
We see that the overall response of these layered networks exhibits an intermediate response in comparison to other uniform networks. 
The response of each node depends strongly on its local environment and the nearby density (see Figure~\ref{fig:schematic} b). 
At frequency $\omega=0.5$, the denser regions have higher responses, while sparser regions of the network have lower responses.
That is, the emergent metamaterial modality of these networks results directly from targeted control of the growth density.

\section{Opportunities for nonequilibrium control}
Nonequilibrium protocol design provides a compelling route to expand our capabilities to manufacture nanoscale metamaterials, especially in contexts where modifying the energetic interactions among components is impossible. 
Here, we take inspiration from reconstitution experiments of dynamically assembling branched actin networks~\cite{li_molecular_2022} and demonstrate that targeting distinct metamaterial properties is possible without altering the underlying components, just by modulating an external control variable with a feedback protocol. 
We also show~\eqref{eq:speedlimit} that the entropy production rate---both in the environment and in an information reservoir---constrains the rate at which we can encode properties into a desired material.
Such fundamental limits shed light on the necessary energetic costs of control, but further investigation of these bounds is required to extract practical design principles for controlling fluctuating materials.

The control framework we introduce here can readily be extended to an experimental setting, even one without mechanisms for feedback. Empirical distributions of control protocols for a specific target can be used as a proxy for a more precise feedback protocols. 
The system we investigate here limits a quantitative investigation into the entropic costs required to encode particular responses because the reference steady state is dissipative. 
However, the computational and theoretical framework motivate a more systematic study into tradeoffs between information acquisition, dissipated heat and the control of a system towards a target state divergent from a nonequilibrium steady state. 
Additionally, this tradeoff can be incorporated into a protocol optimization scheme to design protocols that can encode desired responses with a minimal dissipative cost.

\section*{Acknowledgements}
The authors thank Jiawei Yan for helpful discussions. This material is based upon work supported by the U.S. Department of Energy, Office of Science, Office of Basic Energy Sciences, under Award Number DE-SC0022917. SKM acknowledges the Knut and Alice Wallenberg foundation for financial support through Grant Number KAW 2021.0328.

\paragraph*{Data and Code Availability:}
The data that support the findings of this study are available from the corresponding author upon reasonable request. Our code is available on GitHub at the following url: \href{https://github.com/rotskoff-group/actin-control}{https://github.com/rotskoff-group/actin-control}.

\newpage
\bibliographystyle{unsrt}
\bibliography{actin}

\onecolumngrid
\appendix
\clearpage
\newpage

\setcounter{figure}{0}
\makeatletter 
\renewcommand{\thefigure}{S\@arabic\c@figure}
\makeatother

\section{Minimal model of actin network growth}
\label{app:actin_growth}

We simulate stochastic trajectories of actin network growth under a load using the Gillespie algorithm (Algorithm~\ref{alg:cg}). Under this scheme, we sample reactions according to their relative reaction propensity and sample a reaction time that is inversely proportional to the overall reaction propensity of the system. Finally, after carrying out a reaction, we evolve the position of the barrier.
We define the reaction propensity to be the product of the rate and the number of each of the reactants. For a bimolecular reaction $A + B \to C$, the forward reaction propensity, given a rate $k$, can be written as 
\begin{equation}
    r = kn_An_B,
\end{equation}
where $r$ has units of inverse time.
Given $n$ possible reactions in the system, we select a reaction $j$ to occur based on its relative reaction propensity $\frac{r_j}{\sum_i^{n}{r_i}}$. We define the time of the reaction to be $\tau = \frac{1}{\sum_i^{n}{r_i}} \log\bigl(\frac{1}{u}\bigr)$, where $u$ is a randomly sampled from a uniform distribution defined on $[0, 1].$ 
Finally, consistent with a Brownian ratchet mechanism, we evolve the position of the barrier
according to a Langevin equation, 

\begin{align}
    \dot{h}(t) = F/\gamma + \sqrt{2D} \; \eta(t),
    \label{eq:langevin}
\end{align}
where $\eta(t)$ is a Gaussian random variable with $\avg{\eta(t)}=0$ and $\avg{\eta(t)\eta(t')} = \delta(t-t')$.

\begin{table*}[h]
\centering
\begin{tabular}{|p{10cm}||p{4cm}|}
 \hline
 \multicolumn{2}{|c|}{Reaction Parameters} \\
 \hline
 Parameter & Value\\
\hline
Actin monomer diameter ($\ell_{\rm actin}$) & 2.7 nm  \cite{li_molecular_2022}\\
Capping protein size ($\ell_{\rm cp}$) & 2.7 nm \cite{li_molecular_2022}\\ 
Actin polymerization rate ($k_{\rm actin}^+$) & 10 $\mu$M$^{-1}$s$^{-1}$  \cite{pollard1986rate}\\
Actin depolymerization rate ($k_{\rm actin}^-$) & 1 s$^{-1}$ \cite{pollard1986rate}\\
Arp2/3 binding rate ($k_{\rm Arp2/3}^+$) & 3000 M$^{-1}$s$^{-1}$ \cite{kuhn2007single} \\
Capping binding rate ($k_{\rm cap}^+$) & 2.6 $\mu$M$^{-1}$s$^{-1}$ \cite{kuhn2007single}\\
Actin monomer concentration & 5 $\mu$ M \cite{li_molecular_2022}\\
Capping protein concentration & 100 nM \cite{li_molecular_2022}\\
Arp2/3 concentration & 100 nM \cite{li_molecular_2022}\\
Viscosity of the medium ($\eta$)  & 2.4\; $\eta_{\rm water}$ \cite{zhu2006growth}\\
Viscous drag for a rectangular AFM Cantilever ($\gamma$) & $2 \times 10^{-6}$ kg s$^{-1}$ \cite{liu2010correction}\\
Diffusion constant AFM cantilever\cite{liu2010correction}& $2.07 \times 10^{-15}$ m$^2/s$  \\
Characteristic force scale ($ f_0$)  & 0.75 pN \cite{zhu2006growth}\\

\hline
\end{tabular}
\caption{Experimental parameters used to model growth of actin network}
\label{tab:actin_rates}
\end{table*}

\begin{algorithm}[H]
\caption{Computational model of actin network growwth}
\label{alg:cg}
\begin{algorithmic}[1]
    \State{Initialize system}
    \State{$t = 0$}
    \While {$t < T$}
        \State{Sample reaction $j$ with probability according to relative reaction propensities $\frac{r_j}{\sum_i^{n}{r_i}}$}
        \State{Sample reaction time $\tau =  \frac{1}{\sum_i^{n}{r_i}} \log\bigl(\frac{1}{u}\bigr)$}
        \State{Carry out reaction and update system}
        \State{Evolve barrier height $\dots$}
        \State{$t = t + \tau$}
    \EndWhile
\end{algorithmic}
\end{algorithm}

Furthermore, we impose a reflective boundary condition at the actin network interface, which prevents the barrier from diffusing into the network. We can analytically solve the Fokker-Planck equation corresponding to~\eqref{eq:langevin} with reflective boundary conditions \cite{chandrasekhar1943stochastic,Behringer_hardwall}. For a constant force $F$ and $\tau$, the duration of the reaction, the height of the distribution of the barrier height at the end of the reaction can be described according to 
\begin{equation}
   P(h_{t + \tau}; h_t, \tau) = P_1 (h_{t + \tau}; h_t, \tau) + P_2 (h_{t + \tau}; h_t, \tau) + P_3 (h_{t + \tau}; h_t, \tau).
   \label{eq:fp_reflective}
\end{equation}
Here,
\begin{align}
\begin{split}
    P_1 (h_{t + \tau}; h_t, \tau) &= \frac{1}{\sqrt{4 \pi D \tau}} \exp \left( -\frac{\bigl(h_{t + \tau} - (h_t + \frac{F}{\gamma} \tau)\bigr)^2}{4D \tau}\right),\\
    P_2 (h_{t + \tau}; h_t, \tau) &= \frac{\exp(\frac{-F h_t}{\gamma D})}{\sqrt{4 \pi D \tau}} \exp \left( -\frac{\bigr(h_{t+\tau}+h_t - \frac{F}{\gamma} \tau \bigr)^2}{4D \tau}\right),   \\
   P_3 (h_{t + \tau}; h_t, \tau) &= \frac{-F}{2\gamma D} \exp \left(\frac{F h_t}{\gamma D}\right)  \text{erfc} \left( \frac{\bigl(h_{t+\tau}+h_t +\frac{F}{\gamma} \tau \bigr)}{\sqrt{4D \tau}}\right),
    \end{split}
\end{align}
where erfc denotes the complementary error function. Note that $P_1$ corresponds to the solution of the Fokker-Planck equation with no reflective boundary, while $P_2$ and $P_3$ correspond to transition probabilities that account for a reflective boundary. We can sample a new position for the barrier by numerically solving the equation 

\begin{align}
\begin{split}
    &\int_0^{h_{t + \tau}} dh^\prime\;P_2(h^\prime; h_t, \tau) + P_3(h^\prime; h_t,\tau) \\&= \frac{1}{2} \text{erfc} [\bigl(h_t + \frac{F}{\gamma} \tau\bigr)/\sqrt{4D \tau}] -\frac{1}{2}\exp\left(\frac{Fh}{\gamma D}\right) \;\text{erfc} [\bigl(h_{t+\tau} +h_t + \frac{F}{\gamma} \tau)/\sqrt{4D \tau}]\\& = r
\end{split}
\end{align}
for $h_{t + \tau}$, where $r$ is a uniform random number chosen from the interval [0, $r_{\max}$], where

\begin{align}
    r_{\max} = \int_0^{\infty} dh^\prime\;P_2(h^\prime; h_t, \tau) + P_3(h^\prime; h_t,\tau)
    = \frac{1}{2} \text{erfc} [(h_t + \frac{F}{\gamma} \tau)/\sqrt{4D \tau}].
\end{align}
See Ref.~\cite{Behringer_hardwall} for a detailed discussion and the derivation of the expressions above.

We consider actin network growth on a two-dimensional surface, where only barbed-end growth is allowed. This latter constraint is consistent with the experimental setup in \cite{li_molecular_2022} and is enforced by using profilin-bound actin monomers. Additionally, we fix the chemical potential (i.e.\ the concentration) for all the constituent proteins, allowing us to neglect potential depletion effects. 

We define four possible reactions: actin filament polymerization, actin filament depolymerization, actin filament branching and barbed-end capping, which are detailed below.

\begin{enumerate}
    \item \textbf{Actin filament polymerization}
    \begin{equation}
        r_{\rm{on}} = k_{\rm{actin}}^+ n_{\rm{polymerizable\ filaments}} [\rm{Actin\ Monomers}] 
    \end{equation}
    This reaction results in the addition of an actin monomer to the barbed-end of a single actin filament. Here, $n_{\rm{polymerizable\ filaments}}$ is the number of actin filaments that can be polymerized. Filaments that can be polymerized are those that do not have a capping protein bound and those whose distance between the barbed-end and the barrier is greater than the diameter of an actin monomer ($\ell_{\rm actin}$).
    
    \item\textbf{Actin filament depolymerization}
    \begin{equation}
        r_{\rm off} = k_{\rm actin}^-n_{\rm{depolymerizable\ filaments}}
    \end{equation}
    This reaction results in the removal of an actin monomer from the barbed-end of a single actin filament. Here, $n_{\rm{depolymerizable\ filaments}}$ is the number of actin filaments that can be depolymerized. Filaments that can be depolymerized are those that do not have a capping protein or an Arp2/3 protein bound to the the leading monomer at the barbed-end of the filament.
    
    \item \textbf{Actin filament branching (via Arp2/3)}
    \begin{equation}
        r_{\rm branching} = k_{\rm Arp2/3}^+  n_{\rm{Arp2/3\ binding\ sites}}  n_{\rm{eff\ Arp2/3}}  [\rm{Arp2/3\ Monomers}]
    \end{equation}
    This reaction results in the irreversible binding of the Arp2/3 protein complex to a ``mother'' filament enabling a branched ``daughter'' filament to grow. Biologically, Arp2/3 binding to a mother filament is contingent on a nucleation promotion factor (NPFs). Furthermore, NPFs can be subject to``barbed-end interference'' \cite{li_molecular_2022}, which occurs when uncapped barbed-ends of actin filaments engage with NPFs, restricting Arp2/3 binding.

    For simplicity, we do not explicitly account for NPFs in our model; instead, we implicitly incorporate the effect of NPFs through a scaling factor
    \begin{equation}
        n_{\rm{eff\ arp2/3}} = \max\bigl(0, n_{\rm NPF} - n_{\rm free}\bigr).
    \end{equation}
    Here, $n_{\rm free}$ is the number of free barbed ends within 10$\ell_{\rm actin}$ of the barrier and $n_{\rm NPF}$ represents the number of NPFs, and in practice, is a free parameter that we can tune.

    Finally, $n_{\rm{Arp2/3\ binding\ sites}}$ is the number of potential monomer binding sites for an Arp2/3 protein to bind. These sites include all monomers within 10$\ell_{\rm actin}$ of the barrier that do not already have an Arp2/3 protein bound. Lastly, we note that $k_{\rm Arp2/3\ on}$ is defined as the rate of binding per mother filament, whereas the reaction propensity we compute considers it as the rate of binding per monomer on the mother filament. We absorb this difference into the free parameter $n_{\rm NPF}$.

    \item \textbf{Barbed-end capping }
    \begin{equation}
        r_{\rm capping} = k_{\rm capping}^+ n_{\rm{cappable\ filaments}}[\rm{Capping\ Protein\ Monomers}]
    \end{equation}
    This reaction results in the irreversible binding of a capping protein to the barbed-end of a filament, preventing further actin polymerization and depolymerization. Here, $n_{\rm{cappable\ filaments}}$ corresponds to the number of filaments that can bind a capping protein. Filaments that can be capped are those that do not already have a capping protein bound and those whose distance between the barbed-end and the barrier is greater than the diameter of a capping protein ($\ell_{\rm cp}$).

\end{enumerate}
\section{Minimal model captures key experimental observations}

\begin{figure}[h]
    \centering
    \includegraphics[width=0.495\linewidth]{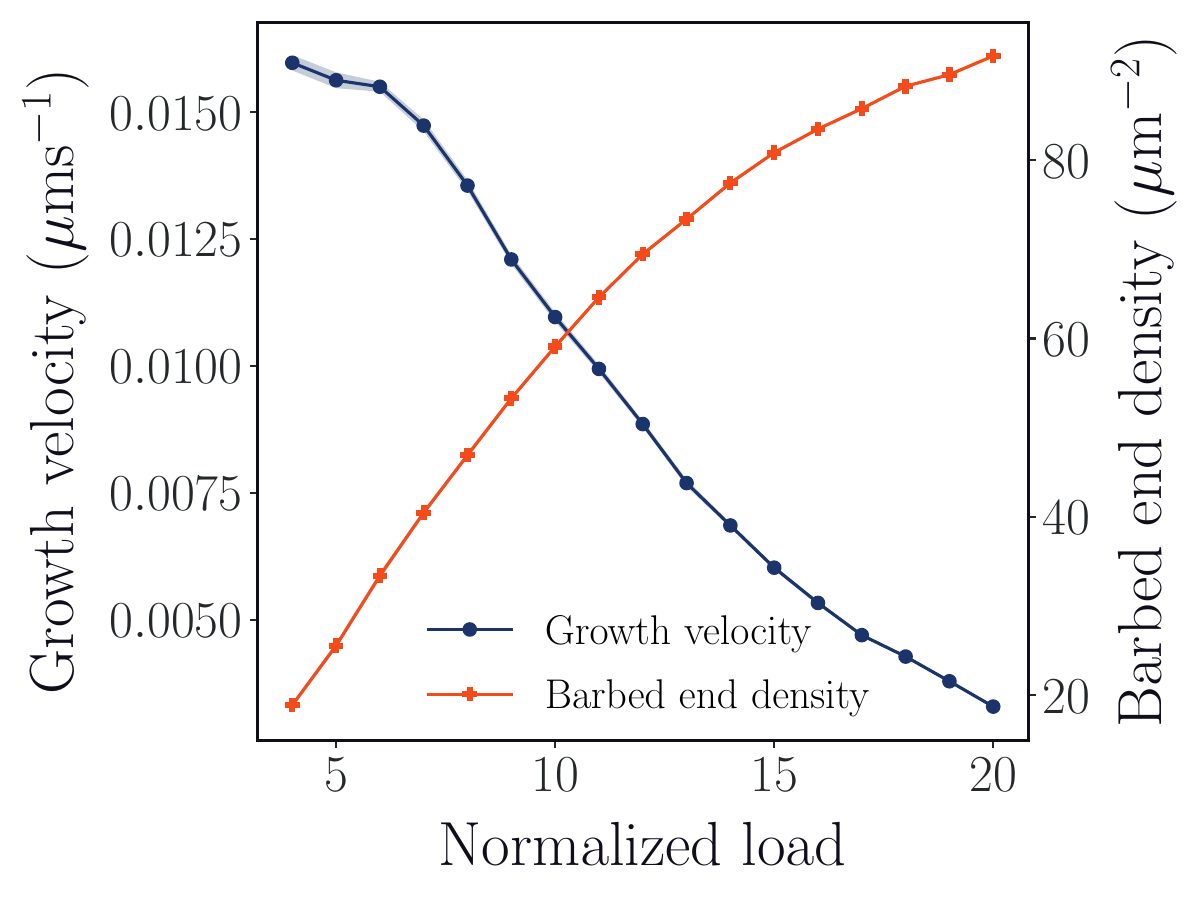}
    \caption{Average growth velocity and barbed end density for different growth loads (normalized by $f_0$) over 500 networks. Under an increased load, actin networks have more free barbed ends and a lower growth velocity. Barbed end density is computed over the region 100 nm from the barrier.}
    \label{fig:vel_be}
\end{figure}

To investigate the validity of our model, we grew an ensemble of networks under a step-wise increasing load (see Figure~\ref{fig:expts} and Figure~\ref{fig:vel_be}) similar to the experimental investigation carried out in \cite{li_molecular_2022}. In these experiments, we initially subject the network to a growth load of $f_0$ (see Table~\ref{tab:actin_rates}) and increase the growth load by $f_0$ every 50 seconds.

First, we observe that under an increased load, the growth velocity of the network decreases (Figure~\ref{fig:expts} a, b). Under an increased load, filament elongation near the barrier is restricted due to steric hindrance, limiting the growth of the network. Second, we observe that under increased load, the density of the actin monomer, capping protein and Arp2/3 within a 100 nm region from the barrier increases (Figure~\ref{fig:expts} c). We consider this restricted region---as opposed to the entire network---to be consistent with experimental measurements obtained via TIRF. Third, we observe that the propensities for branching and capping are balanced (Figure~\ref{fig:expts} d) across different loads. Importantly, the propensity for branching decreases with increasing load as a result of barbed-end interference, while the propensity for capping decreases with increasing load as a result of steric hindrance with the barrier. This balance in propensities between capping and branching ensures steady-state network growth. Finally, we observe an inverse relationship between the growth velocity and the free barbed-end density (Figure~\ref{fig:vel_be}). Increases in the barbed-end density interfere with branching, which in turn inhibits network growth and causes a reduction in the growth velocity. Furthermore, we capture the phenomenon that subsequent increases in the barbed end density decreases with increasing load, leading to a decreasing slope in the curve.  

Together, these results demonstrate that our model captures most of the key observations from experiments, making it a suitable environment for investigating control of actin network growth.

\section{Structural mechanics of actin networks}
\label{app:struc_mech}
\begin{figure*}[h]
    \centering
     \includegraphics[width=\linewidth]{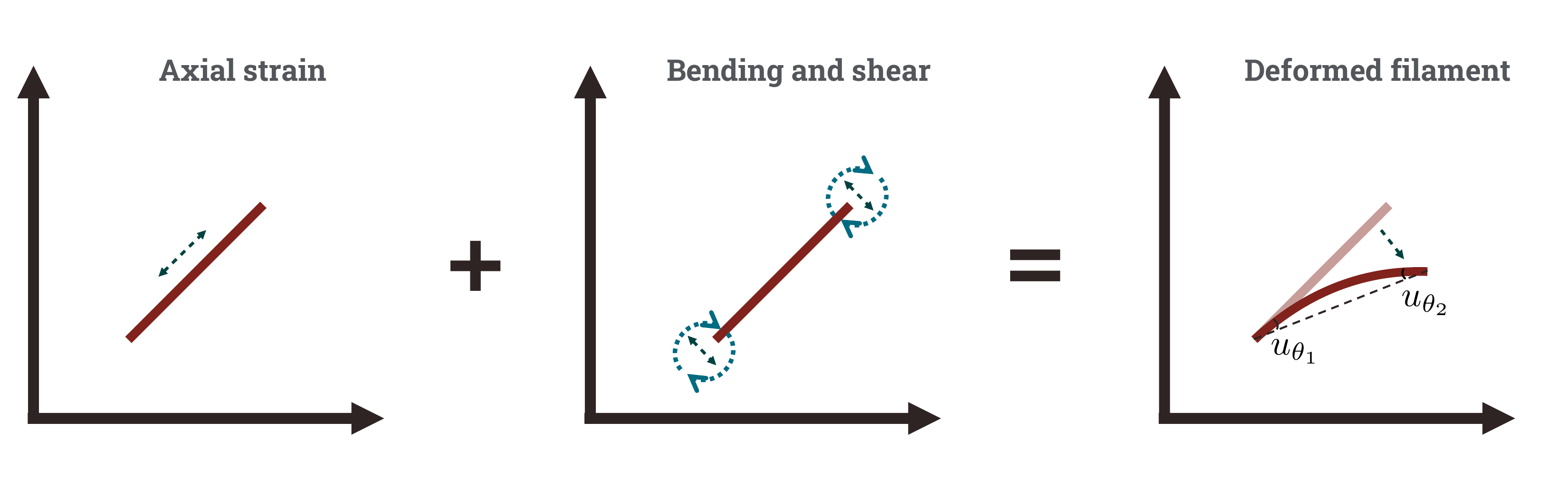}
    \caption{Schematic of actin filament deformations. Axial strains deform filament along principal axis, bending strain deforms filament away from the principal axis, and shear strain deforms filament's cross section.}
    \label{fig:bending_schematic}
\end{figure*}
We investigate the response of actin networks to external compressive forces in order to gain insight into the elastic properties of networks grown under different load forces. During the growth process, outside of daughter filaments that branch from a mother filament, filaments within the network are not interconnected. To investigate network-level mechanical responses to compressive forces, we consider intersecting filaments to be crosslinked---this can be enforced in an experiment setting via a crosslinking protein such as filamin \cite{bieling_force_2016}. Ultimately, this framework allows us to consider the actin network as a collection of nodes and edges, where nodes correspond to either a branching point, a crosslinking point or the end of a network. 

We consider filament (i.e. edge) deformations that include an axial strain, a bending strain and a shear strain (Figure~\ref{fig:bending_schematic}). The axial strain corresponds to compressions or elongations of the filament along the principal axis of the edge, the bending strain corresponds to curvatures along the filament away from the principal axis, and the shear strain corresponds to distortion of the filament's cross-section. A filament experiences a combination of these deformations simultaneously under a compressive load, and we can compute this overall deformation by defining a stiffness matrix that we use in conjunction with Hooke's law to determine equilibrium network responses to compressive
forces.

More precisely, given a compressive force $\Fb$ and a stiffness matrix $K$, we can compute the deformations of each edge 
\begin{equation}
        \ub = K^{-1}\Fb,
        \label{eq:hookes_law}
\end{equation}
where $\ub = [u_{x_1}, u_{y_1}, u_{\theta_1}, u_{x_2}, u_{x_2}, u_{\theta_2}]^T$. Here, $u_{x_i}$ and $u_{y_i}$represents the change in the $x$ and $y$ coordinates of the $i$th node and $u_{\theta_i}$ corresponds to the deviation of the $i$th node from the principal axis (see Figure~\ref{fig:bending_schematic}). 

We define the compressive force $\Fb = [F_{x_1}, F_{y_1}, F_{M_1}, F_{x_2}, F_{y_2}, F_{M_2}]$, where $F_{x_i}$ and $F_{y_i}$ are the forces on the $x$ and $y$ directions of the $i$th node and $F_{M_i}$ is the torque acting on the $i$th node.

For an actin filament with cross-sectional area $A$, length $L$, modulus elasticity $E$, moment of inertia $I$, and unit vector $\eb$, the stiffness matrix\cite{mcguire1982matrix,bathe2006finite}, is given by 
\begin{equation}
        K=\frac{E}{L}\scalebox{0.9}{$\begin{bmatrix}
                Ae_x^2 +\frac{12Ie_y^2}{L^2} & \left(A-\frac{12I}{L^2} \right)e_xe_y & -\frac{6I}{L}e_y &-\left( Ae_x^2+\frac{12I}{L^2}e_y^2\right) &-\left(A-\frac{12I}{L^2} \right)e_xe_y &-\frac{6I}{L}e_y  \\
                & Ae_y^2 +\frac{12I}{L^2}e_x^2 & \frac{6I}{L}e_x &-\left(A-\frac{12I}{L^2} \right)e_xe_y &-\left(Ae_y^2 +\frac{12Ie_x^2}{L^2} \right)&\frac{6I}{L}e_x \\
                &  & 4I &\frac{6I}{L}e_y &-\frac{6I}{L}e_x &2I  \\
                &  &  &Ae_x^2 +\frac{12Ie_y^2}{L^2}  &\left(A-\frac{12I}{L^2} \right)e_xe_y &\frac{6I}{L}e_y  \\
                & \rm{sym}  &  & &Ae_y^2 +\frac{12I}{L^2}e_x^2&-\frac{6I}{L}e_x  \\
                &  &  & & &4I 
                \end{bmatrix}$}.
    \label{eq:sm_edge}
\end{equation}
We can compute network-level responses to a compressive force by defining a stiffness matrix for a network of $N$ nodes and $M$ edges. First, we lift the edge-level stiffness matrix $K \in \mathbb{R}^{6\times6}$ to the stiffness matrix $\mathcal{K} \in \mathbb{R}^{3N\times3N}.$ 

Given an edge $i$ that connects nodes $a$ and $b,$ and $r,s \in \lbrace 0, 1,2 \rbrace,$ we define 
\begin{equation}
    \mathcal{K}^i_{\alpha,\beta} = 
\begin{cases}
    K^i_{r,s} ,& \text{if } (\alpha, \beta) = (3a+r,3a+s) \\
    K^i_{r,3+s},& \text{if } (\alpha, \beta) =(3a+r,3b+s) \\
    K^i_{3+r,s} ,&\text{if }(\alpha, \beta) =(3b+r,3a+s) \\
    K^i_{3+r,3+s} ,&\text{if } (\alpha, \beta) =(3b+r,3b+s) \\
    0,              & \text{otherwise.}
\end{cases}
\label{eq:stiff_lift}
\end{equation}
Conceptually, in \eqref{eq:stiff_lift}, we take each quadrant of $K^i$ and fill in the appropriate region of $\mathcal{K}^i$ corresponding to node $a$ and $b$.

We can then compute the global stiffness matrix of the network 
\begin{equation}
    \mathcal{K}_{\rm global} = \sum_{i = 1}^M \mathcal{K}^i.
    \label{eq:glob_stiff}
\end{equation}

Finally, to realize a physically meaningful compression, we enforce a rigid base of the network by fixing all nodes within a cutoff of $0.5\ \mu$m from $y=0$. We remove the rows and columns of the stiffness matrix $\mathcal{K}$ that corresponds to these fixed nodes and can use \eqref{eq:hookes_law} to compute the deformation of every filament in the network.

\subsubsection*{Stress-strain response of the network}

\begin{figure}
    \centering
    \includegraphics[width=0.495\linewidth]{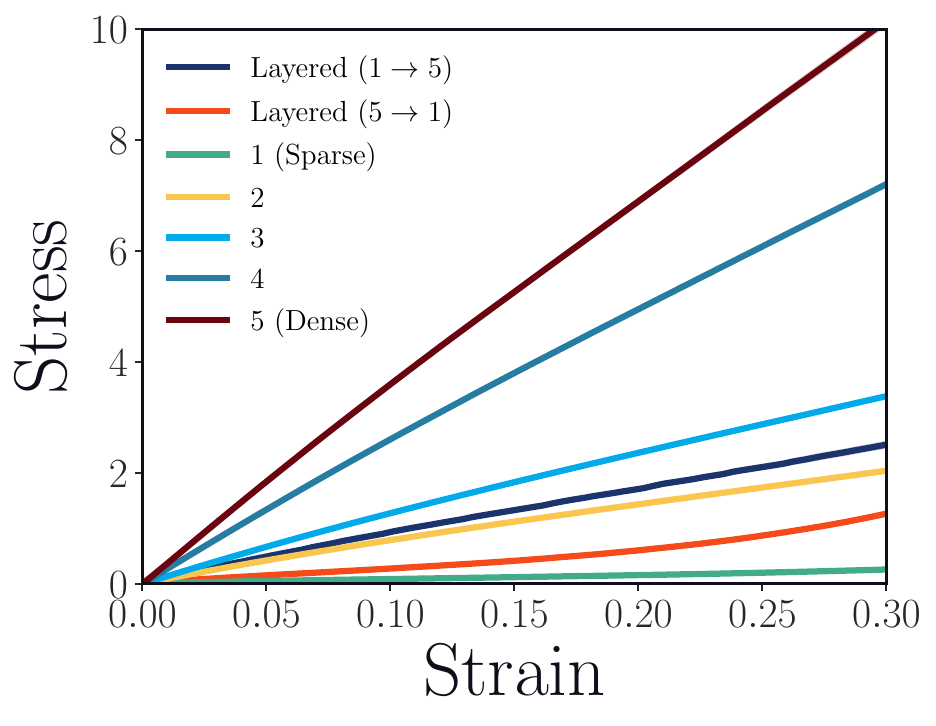}
    \caption{Plot of Stress ($\Sigma$) vs. Strain ($\Delta$) for 200 networks of different composition. See Figure~\ref{fig:response} caption for description of legends. The strain represents the fraction of compression of the network and stress represents the applied force required to achieve the strain }
    \label{fig:stress_strain}
\end{figure}

\begin{algorithm}[H]
\caption{Stress-strain procedure}
\label{alg:stress_strain}
\begin{algorithmic}[1]
    \State{$\Delta_0 = 0$}
    \State{$\sigma_0 = 0$}
    \State{$i = 1$}
    \While {$\Delta_i < \Delta_{\rm thresh}$}
        \State{Apply compressive force $F^{\rm stress}_i = iF^{\rm stress}_0$ according to \eqref{eq:hookes_law}}
        \State{$\Delta_i = (h_{0} - h_i / h_0)$}
        \State{$\sigma_i = iF^{\rm stress}_0 \tilde{N}$}
        \State{$i = i + 1$}
    \EndWhile
\end{algorithmic}
\end{algorithm}

We quantify the mechanical response of the network by measuring the stress (i.e.\ deformation) of the network under an external strain (i.e.\ external force). For simplicity, we work with units where the axial rigidity $AE = 1$ and the ratio $EI/AE = 0.01$.  

To obtain the stress-strain profile, we apply a series of linearly increasing compressive forces on the network and compute the strain at step $i$ as 
\begin{equation}
    \Delta_i = \frac{h_0 - h_i}{h_0}
    \label{eq:strain}
\end{equation}
and the stress at step $i$ as 
\begin{equation}
    \Sigma_i = \sum_i \sigma_i,
    \label{eq:stress}
\end{equation}
where $\sigma_i = F^{\rm stress}_i \tilde{N}_i$, $F^{\rm stress}_i = iF^{\rm stress}_0$ and $\tilde{N}_i$ is the number of nodes within 0.2 $\mu$m of the growth-front of the network. We work with small compressions and define $F^{\rm stress}_0 = 0.001$. Here, the stress is measured in units of $AE$.

In the curves shown in Figure~\ref{fig:stress_strain}, we set $\Delta_{\rm thresh} = 0.3$ resulting in compression of the network to $70\%$ of its initial height. We see that uniform networks with higher density have increasingly stiffer responses, while layered networks have intermediate responses. Because we're only applying a stress on the top region of the network, the layered network with a denser growth front ($1\to 5$) has a stiffer response in comparison to the layered network with a sparser growth front ($5\to 1$).

\section{Response of actin networks under time periodic forces}
\label{app:time_periodic}
We apply a time-periodic force to actin networks to study the frequency response of networks grown under different load conditions. For a time-periodic force $\Fb(t) = \Fb_0 \exp(i\omega t)$ applied to a network, the motion of the nodes in a single edge evolve according to
\begin{equation}
    M\Ddot{\ub} = K {\ub}  + {\Fb}(t), 
    \label{eq:time_response}
\end{equation}
where $M$ is the mass matrix that accounts for the nonzero mass of an edge. For a filament (i.e.\ edge) with mass density $\rho,$ length $L$ and unit vector $\eb,$ the mass matrix \cite{rao2017finite,mcguire1982matrix} is given by
\begin{equation}
    M =\frac{\rho L}{420} \scalebox{0.8}{$
\begin{bmatrix}
 4 \left(35 e_x^2+39 e_y^2\right) & -16 e_x e_y & -22 L e_y & 70 e_x^2+54 e_y^2 & 16 e_x e_y & 13 L e_y \\
  & 4 \left(39 e_x^2+35 e_y^2\right) & 22 e_x L & 16 e_x e_y & 54 e_x^2+70 e_y^2 & -13 e_x L \\
 &  & 4 L^2 & -13 L e_y & 13 e_x L & -3 L^2 \\
& & & 4 \left(35 e_x^2+39 e_y^2\right) & -16 e_x e_y & 22 L e_y \\
&  & \rm{sym} &  & 4 \left(39 e_x^2+35 e_y^2\right) & -22 e_x L \\
 &  &  &  &  & 4 L^2
\end{bmatrix}$}
\end{equation}

As with the stiffness matrix in Section~\ref{app:struc_mech}, we can compute a mass matrix for the entire network.

Given edge $i$ that connects nodes $a$ and $b,$ and $r,s \in \lbrace 0, 1,2 \rbrace,$ we define 
\begin{equation}
    \mathcal{M}^i_{\alpha,\beta} = 
\begin{cases}
    M^i_{r,s} ,& \text{if } (\alpha, \beta) = (3a+r,3a+s) \\
    M^i_{r,3+s},& \text{if } (\alpha, \beta) =(3a+r,3b+s) \\
    M^i_{3+r,s} ,&\text{if }(\alpha, \beta) =(3b+r,3a+s) \\
    M^i_{3+r,3+s} ,&\text{if } (\alpha, \beta) =(3b+r,3b+s) \\
    0,              & \text{otherwise.}
\end{cases}
\label{eq:mass_lift}
\end{equation}

We can then compute the global mass matrix of the network 
\begin{equation}
    \mathcal{M}_{\rm global} = \sum_{i = 1}^M \mathcal{M}^i.
    \label{eq:glob_mass}
\end{equation}
In Figure~\ref{fig:response} b), we compute the eigenvalues of the dynamical matrix $\mathcal{D}_{\rm global} = \mathcal{M}_{\rm global}^{-1}\mathcal{K}_{\rm global}$.

Using $\mathcal{M}_{\rm global}$ and the global stiffness matrix $\mathcal{K}_{\rm global},$ we can use \eqref{eq:time_response} to solve for the response of the entire network.

The solution can be written down as
\begin{equation}
    {\ub}(t)  ={\ub}_0 \exp(i\omega t),
    \label{eq:time_response_sol}
\end{equation}
where ${\ub}_0 = \mathcal{G}(\omega){\Fb}_0$ and $\mathcal{G}(\omega) = (-\omega^2 \mathcal{M}_{\rm global} + \mathcal{K}_{\rm global})^{-1}.$

For a periodic force with frequency $\omega,$ we can compute the time-averaged covariance matrix of node responses
\begin{align}
\begin{split}
    C(\Fb_0, \omega) &=  \frac{1}{2\pi \omega} \int_0^{\frac{2\pi}{\omega}} \text{Re}(\ub) \text{Re}(\ub)^T dt \\
                     &= \mathcal{G}(\omega) {\Fb}_0 {\Fb}_0^\dag \mathcal{G}^\dag(\omega),
                     \end{split}
    \label{eq:time_avg_responses}
\end{align}
where $(\cdot)^\dag$ denotes a conjugate transpose.

Given $\Fb_0,$ we can compute the response of the network as
\begin{equation}
    R_{\Fb_0}(\omega )= \text{Tr} \bigl( C(\Fb_0, \omega) \bigr),
    \label{eq:force_repsonse}
\end{equation}
where $\text{Tr}((\cdot))$ denotes the trace. 

Finally, for an ensemble of forces that satisfy $\langle {\Fb}_0{\Fb}_0^{\dag}\rangle = \mathbf{I},$ we can compute an average response 
 \begin{align}
\begin{split}
        R(\omega)&= \langle R_{{\Fb}_0}(\omega ) \rangle_{{\Fb}_0} \\
        &= \text{Tr} \left( {\mathcal{G}}(\omega) \langle {\Fb}_0 {\bm F}_0^\dag \rangle {\mathcal{G}}^\dag(\omega) \right) \\
        &= \text{Tr} \left( {\mathcal{G}}(\omega) {\mathcal{G}}^\dag(\omega) \right).
\end{split}
\label{eq:avg_repsonse}
\end{align}

The response will be maximum at frequencies where $\omega^2$ is an eigenvalue of the stiffness matrix as ${\mathcal{G}}^{-1}(\omega)$ yields a null matrix at those frequencies.

\section{Computational details of control framework}
\label{app:rl}
\begin{algorithm}[H]
\caption{Clipped Double Q-Learning Training}
\label{alg:train_cdql}
\begin{algorithmic}[1]
    \State{Initialize Replay Buffer $\mathcal{R}$}
    \State {Randomly initialize $Q_1$ and $Q_2$ networks with weights ${\theta^{Q_1}}$ and ${\theta^{Q_2}}$}
    \State {Initialize target networks $Q_1^{\prime}$ and $Q_2^{\prime}$ networks with weights ${\theta^{Q_1^{\prime}} \leftarrow \theta^{Q_1}}$ and ${\theta^{Q_2^{\prime}} \leftarrow \theta^{Q_2}}$}
    \For {e = $0$ ${\dots}$ M}
    \State{Grow network for time $\tau_{\rm burn\ in}$}
        \For {t = $0$ ${\dots}$ T}
            \If {e ${< e_{\rm explore}}$}
                \State{Select a random action ${a_t}$ from $\mathcal{A}$}
            \Else
                \State{Select ${a_t = \argmin_{a} Q_1(\sb_t, a)}$}
            \EndIf
         \State{Execute action ${a_t}$ and observe next state ${\sb_{t+1}}$ and reward $r_t$}
         \State{Store transition ${(\sb_t, a_t, r_t, \sb_{t+1})}$ in $\mathcal{R}$}

         \State{Sample random batch of transitions ${(\sb_j, a_j, r_j, \sb_{j+1})}$ from $\mathcal{R}$}
         \State{Set ${y_j = r_j + \gamma \max_{1, 2}\min_{a^{\prime}} Q_{1, 2}^{\prime}(\sb_{j + 1}, a)}$}

        \State{Update ${Q_{1, 2}}$ by minimizing ${L_{1, 2} = (y_j - {Q_{1, 2}(\sb_j, a_j)})^{2}}$}
        \State{Update the target networks ${\theta^{Q_{1, 2}^{\prime}} \leftarrow (1 - \tau){\theta^{Q_{1, 2}^{\prime}}} + \tau{\theta^{Q_{1, 2}}}}$}
        \EndFor
    \EndFor
\end{algorithmic}
\end{algorithm}

\begin{algorithm}[H]
\caption{Controlled actin network growth}
\label{alg:grow_network}
\begin{algorithmic}[1]
    \State{Load trained value functions $\{Q^i\}_{i=1}^n$}
    \State{Initialize system}
    \State{$t = 0$}
    \While {$t < T$}
        \State{Update load force $F = u(\sb_t, t)$}
        \State{Simulate system for 1 second}
        \State{$t=t+1$}
    \EndWhile
\end{algorithmic}
\end{algorithm}

\begin{table*}
\centering
\begin{tabular}{|p{4cm}||p{8cm}|}
 \hline
 \multicolumn{2}{|c|}{Hyperparameters} \\
 \hline
 Name & Value\\
 \hline
 $Q$ learning rate  & $3 \times 10^{-4}$\\
 Optimizer&   Adam\\
 Target Update Rate ($\tau$) & $5 \times 10^{-3}$\\
 Batch Size    &32\\
 Discount Factor ($\gamma$)&   0.9 \\
 Number of Hidden Layers & 2\\
 Network Width & 64\\
 $T$ & 50  \\
 $\mathcal{A}$ (Normalized by $f_0$) & $[1.0, 2.0, 4.0, 6.0, 8.0, 10.0, 12.0, 14.0, 16.0, 18.0]$\\
 \hline
\end{tabular}
\caption{Relevant parameters for $Q$-learning training}
\label{tab:dql_hyp}
\end{table*}

We employ a model-free, off-policy Reinforcement Learning (RL) based on Deep Q-Learning to target spatial density profiles for actin networks. We use a variant of Q-Learning, known as clipped double Q-learning \cite{mnih-atari-2013, chennakesavalu_probing_2021, chennakesavalu_cooperative_2021}, that incorporates two Q networks to avoid an underestimation bias and uses a soft policy update \cite{fujimoto_td3_18} to ensure a more stable training process. 

Generally, with reinforcement learning, we are interested in training an agent to optimize interactions (i.e.\ actions) with an uncertain environment in order to optimize some reward \cite{sutton_reinforcement_2018}. Here, our environment consists of an actin network growing against a load and the agent is an external controller that can take action on the environment by modulating the strength of the load force.

Because the capping protein arrests filament elongation and shrinkage and because filaments that are not proximal to the barrier are quickly capped, in practice, we can only reasonably control the \textit{growth-front} of the actin network. We define the growth-front of the network to be the region within $10\ell_{\rm actin}$ from the barrier where Arp2/3 binding---and as a result branching---can take place. Given a task of targeting a growth-front density of $f_*^i$, we define the task-specific reward function 
\begin{equation}
    R^i(\xb) = \bigl|f_{\rm gf}(\xb) - f_*^i\bigr|.
\end{equation}
We note that in this setting, we aim to minimize the reward in order to reliably target specific densities.

With Q-learning, we learn a state-action value function 
\begin{equation}
\label{eq:Q}
    Q(\xb, \ab) = \EE_u \left[\sum_{k=0}^\infty \gamma^k R(\xb_{t+1}) \right],
\end{equation}
where $Q(\xb, a)$ represents the expected value of a discounted cumulative reward upon taking a particular action $a$ given a current state $\xb$ with subsequent actions according to a protoocl $u$. We consider a discrete action space $\mathcal{A}$ (see Table~\ref{tab:dql_hyp}) to limit potential experimental challenges of applying load forces with arbitrary resolution. 

In practice, we will not have access to the full state of the system $\xb$; to account for the limited  resolution we can access experimentally, we define the state of the system $\sb \in \mathbb{R}^3$, to be the actin monomer density, ARP2/3 density and the capping protein density within 100 nm of the barrier. These densities are directly comparable to the experimental fluorescent measurements of actin monomer, capping protein and Arp2/3 obtained via TIRF.

We are interested in targeting a set of growth-front densities $\{f_i^*\}_{i=1}^n,$ to both grow networks with various densities and to create layered networks. We approach this with a multi-task RL framework, where given a target growth-front density of $f_*^i$ and the corresponding reward function $R^i$, we train a state-action value function $Q^i(\sb, a)$ (see Algorithm~\ref{alg:train_cdql}). For each task $i$, we use the trained $Q_1$ as our $Q^i$.
In our framework, each $Q^i$ is an individual neural network, but, in practice, could also be a single neural network $Q(\cdot, i)$, with a task label. 

Given a set of trained value functions, $\{Q^i\}_{i=1}^n$, our control protocol is
\begin{equation}
    u(\sb_t, t) = \argmin_{a \in \mathcal{A}}Q^{i(t)}(\sb_t, a),
    \label{eq:ctrol}
\end{equation}
where $i(t)$, represents the index of the desired growth-front density at some time $t$. We manually define $i(t)$ to grow layered networks or homogeneous networks of different densities (see Algorithm~\ref{alg:grow_network}).

\begin{figure}[h]
    \centering
    \includegraphics[width=0.995\linewidth]{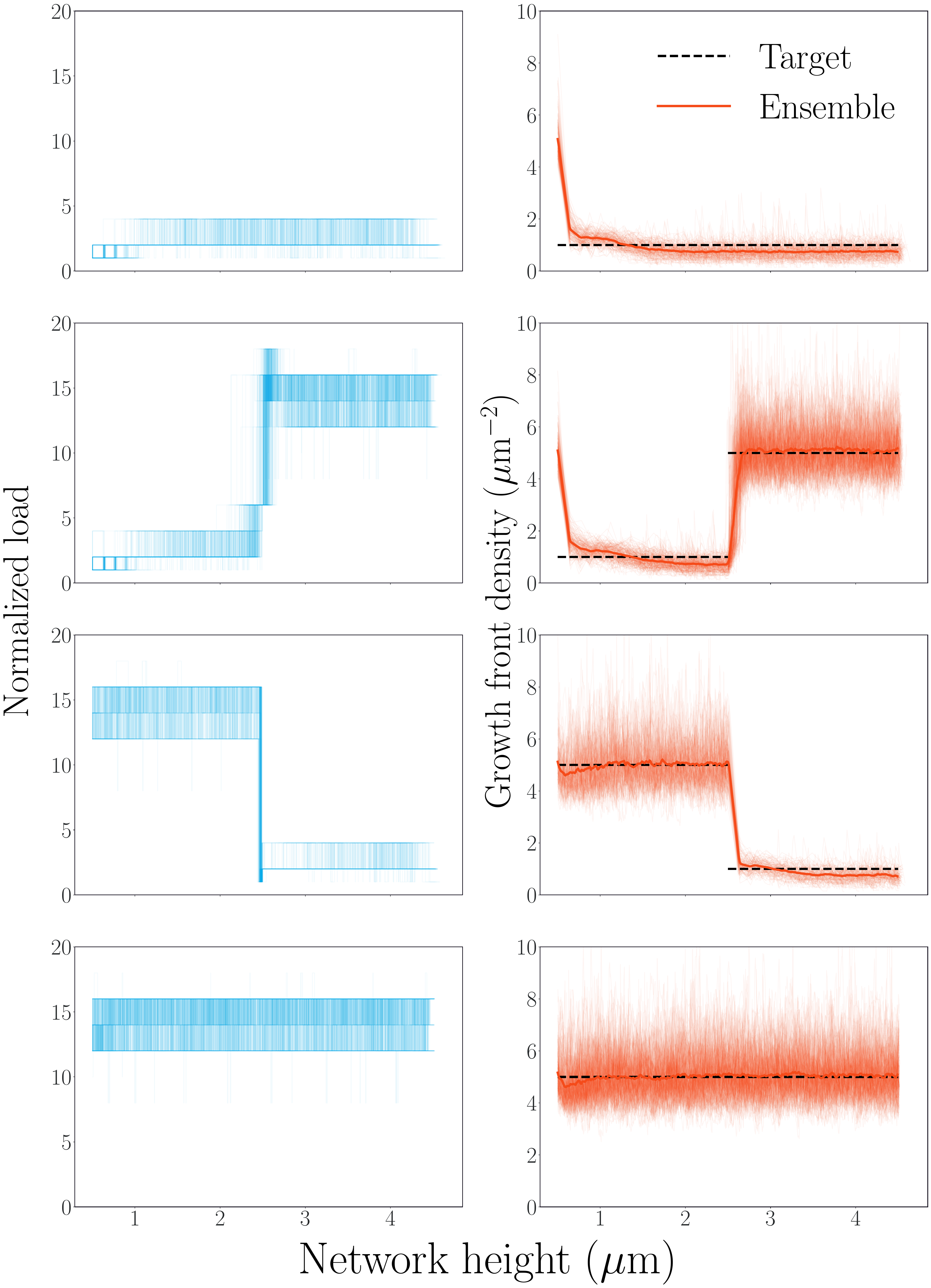}
    \caption{Ensemble of 200 networks grown to a target layered structure with state dependent protocols. 
    From top to bottom, target networks are sparse network, layered network with soft base, layered network with stiff base and dense network. The ensemble of state-dependent control protocols (left) enable realization of target growth-front density (right).}
    \label{fig:rl_all}
\end{figure}

\begin{figure}
    \centering
    \includegraphics[width=\linewidth]{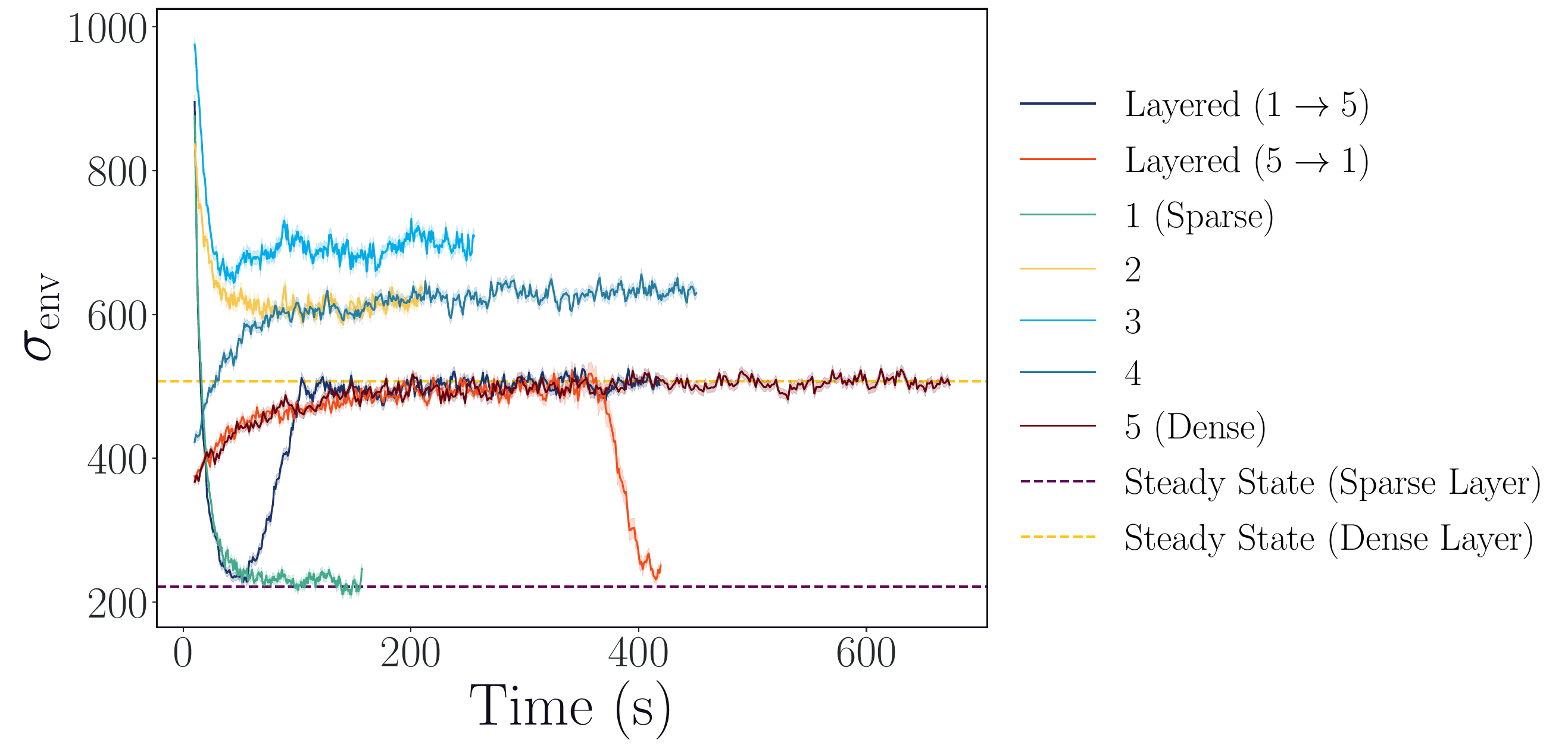}
    \caption{The entropy production rate associated with actin filament growth during feedback controlled growth. Transients in the growth dynamics differ markedly from the steady state behavior. In all cases, the total entropy production is positive, demonstrating that the information reservoir associated with the protocol is being utilized to do work on the system.}
    \label{fig:epr}
\end{figure}

\end{document}